\documentclass{elsart}

 \usepackage{graphicx}

\usepackage{amssymb}

\begin{document}

\begin{frontmatter}

\title{Hard Probes at LHC}

% use optional labels to link authors explicitly to addresses:
% \author[label1,label2]{}
% \address[label1]{}
% \address[label2]{}

\author{Berndt M\"uller}

\address{Department of Physics, Duke University, Durham, NC 27708, USA}
\address{YITP, Kyoto University, Kyoto 606-8502, Japan} 

\begin{abstract}
This talk gives a theoretical perspective of the physics issues 
awaiting us when heavy ions will collide in the LHC.
\end{abstract}
\end{frontmatter}

% main text

\section{Introduction}
\label{sec:intro}

The heavy ion physics program at the Large Hadron Collider (LHC) at
CERN will commence in about two years from now. It is thus appropriate
to ask what we can expect from heavy ion collisions at the LHC, what 
our physics goals are or should be, and whether the theory community 
is adequately prepared for the anticipated scientific challenges. 
In other words, what is our grand strategy for the LHC heavy ion 
program? I will try to give answers to these questions in my lecture.

It is, first of all, important to recognize that the LHC program will
constitute the last step of relativistic heavy ion physics into 
completely uncharted territory for a long time to come, maybe forever.
There is not another accelerator on the planning horizon which could
provide higher collision energies, and we have almost (with the exception
of $^{238}$U) exhausted the range of accessible nuclear masses. It is also 
important to keep in mind that the increase of center-of-mass energy from 
RHIC to LHC is larger, measured on a logarithmic scale, that the step up 
from SPS to RHIC.

On the other hand, the data taken at RHIC have provided clear evidence 
that energy densities solidly, maybe even far, in excess of the critical
energy density of QCD (about 1 GeV/fm$^3$) have been reached there, 
implying that the transition from hadronic matter to quark-gluon plasma
has been made. According to common wisdom, higher beam energies will 
only provide for a hotter quark-gluon plasma, but not something entirely 
new. This argument has been thrown into doubt in recent years, when 
experimental evidence from RHIC mounted showing that the produced 
matter is highly fluid and opaque to probes with open color. If the 
matter discovered at RHIC is a ``strongly coupled'' quark-gluon plasma,
then maybe the matter that will be produced at LHC energies has a very
different structure, more that of a ``weakly coupled'' gaseous plasma.

It must be noted, however, that lattice gauge calculations do not show 
evidence for a second transition point above $T_c$, where the structure
of QCD matter might change in a qualitative way. The lattice results 
show the effects of interactions to gradually diminish as the temperature
rises, due to the logarithmic weakening of $\alpha_s(T)$, but there is 
no sign of a dramatic change in thermodynamic quantities or correlation
lengths once the transition from hadron gas to quark-gluon plasma is 
complete. Speculations that the matter produced at LHC will be qualitatively
different from that observed at RHIC thus lack a solid theoretical basis.

Taking a conservative attitude, one is led to the expectation that LHC 
will produce a similar, but hotter type of matter as RHIC. Hard QCD 
phenomena, which were accessible at RHIC for the first time, will be 
much more abundant and thus can be more easily studied experimentally.
The central question for theorists will, therefore, be whether the 
theoretical framework that has been developed at RHIC will hold up 
when the data from LHC experiments come in. The extended kinematic 
range offered by the LHC will provide for quantitative test of the 
models and concepts that have successfully described the RHIC data:
\begin{itemize}
\item The saturation of initial parton densities reflected in the 
      rapidity distribution of the charged particle multiplicity;
\item The almost ideal hydrodynamical evolution of the matter as 
      evidenced in the magnitude of the elliptic flow ($v_2$);
\item The scaling of parton energy loss with the path weighted, 
      integrated matter density $\int \rho\tau d\tau$;
\item Color screening on subhadronic scales leading to bulk hadronization
      and hadron formation by valence quark recombination;
\end{itemize}
The major new probes of matter accessible at LHC energies -- $b$-quarks
and resolved jets, will permit tests of these theoretical ideas with
much improved control on the theoretical predictions (because of the
larger momentum scales involved).

If the general picture developed and still being refined at RHIC is 
found to apply to the LHC data, as well, we will be able to proclaim 
success. Success, that is, in having obtained a theoretically firmly
grounded understanding of the properties of hot QCD matter -- the 
quark-gluon plasma -- and of the space-time evolution of relativistic
heavy ion collisions at the highest energies. This would be a major 
achievement for our field. On the other hand, if the LHC experiments 
were to bring unforeseen surprises showing, for example, that the matter 
created at the higher energies is qualitatively different from that
formed at RHIC, this would be most exciting, as well. Some would 
probably find such a scenario more exciting, because it is the tendency
of scientists (and funding agencies!) to get easily bored by repeated 
confirmations of previously made discoveries.

The questions we turn to next are: What do we currently expect nuclear
collisions at LHC to look like? Are we ready to make quantitative 
predictions for LHC data on the basis of our present framework?

\section{Expectations for ``LH-I-C''}

\begin{figure}
\centerline{\includegraphics[width=0.7\linewidth]{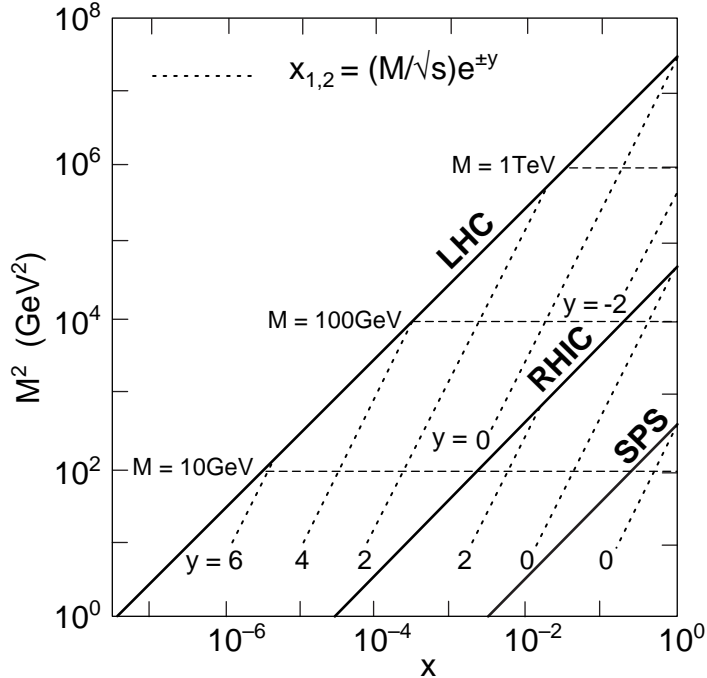}}
\caption{Parton kinematic range of the LHC for Pb+Pb at the highest
         energy ($\sqrt{s_{\rm NN}}=5.5$ TeV). The ranges accessible
         at the top RHIC and SPS energies are shown in comparison.
         From ref.~\cite{Vogt:2004hc}.}
\label{fig:kin-range}
\end{figure}

At a maximal nucleon-nucleon center-of-mass energy $\sqrt{s_{\rm NN}}=5.5$ TeV
for Pb+Pb, compared with $\sqrt{s_{\rm NN}}=200$ GeV for RHIC, the LH-I-C
(the LHC operated with heavy ions) will have a much larger kinematic 
range than RHIC (see Fig.~\ref{fig:kin-range})
\cite{Vogt:2004hc,Muller:2004dd}. Jets with $E_T=100$ GeV 
at LH-I-C will probe the same parton structure of the colliding nuclei 
(measured in terms of Bjorken-$x$) as 2 GeV leading hadrons at RHIC. 
The initial conditions for soft physics at mid-rapidity at LH-I-C 
will be determined by the parton structure of nuclei in the range 
$10^{-4}<x<10^{-3}$, as opposed to $x\approx 10{-2}$ at RHIC. The far 
forward and backward regions will provide access to the parton
distributions of nuclei down to $x\approx 10^{-5}$.

Let us begin with the expectations for the environment that will be
probed by hard processes in nuclear collisions at the LHC. What kind
of matter do we expect these collisions to produce at mid-rapidity? 
The physical picture of the initial state corresponding to the parton 
structure of a $^{208}$Pb nucleus at $x\leq 10^{-3}$ is that of a 
saturated gluon distribution \cite{McLerran:1993ni,McLerran:1993ka}.
At low scale $Q^2\leq Q_s^2$, small $x$, and for sufficiently large 
nuclei $A$, the gluon density in the nuclear parton distribution 
becomes so large that nonlinear QCD interactions lead to its saturation.
The saturation scale $Q_s$ at a given $x$ and transverse position 
${\mathbf b}$ is given by the relation \cite{Mueller:2002kw}
\begin{equation}
Q_s^2(x,{\mathbf b}) \approx \frac{4\pi^2\alpha(Q_s^2)N_c}{N_c^2-1}
  xG_N(x,Q_s^2)T_A({\mathbf b}),
\label{eq:Qsat}
\end{equation}
where $G_N$ is the gluon distribution in the nucleon and $T_A$ is 
the longitudinally integrated nuclear density profile. The numerical
value applicable to the LHC is not precisely known, $Q_s\approx 2$ GeV/c
is a reasonable estimate \cite{Gyulassy:1997vt,Eskola:1999fc,Lappi:2006hq}.
It certainly would be good to obtain a more reliable determination of
$Q_s$ for RHIC and use it to extrapolate into the LHC energy range.

When the two nuclei collide, the saturated gluons are scattered on-shell
and released \cite{Kovchegov:2000hz}, creating a state sometimes referred 
to as the {\em glasma} \cite{Lappi:2006fp}, which can be described by 
semiclassical color fields. As has been shown recently, this form of 
unequilibrated matter almost immediately equilibrates chemically into 
a quasi-thermal {\em quark-gluon plasma} \cite{Gelis:2005pb,Lappi:2006nx}.
Using these concepts it is possible to predict the charged particle
multiplicity and transverse energy per unit rapidity expected in Pb+Pb
collisions at the LHC. Such predictions have been made using schematic
saturation models \cite{Eskola:2001bf,Armesto:2004ud}, as well as in the
framework of classical color field dynamics \cite{Krasnitz:2003jw}. A 
third approach uses $k_T$-factorization to calculate the release of 
gluons from a saturated distribution in the fragmentation region and 
and extrapolates to mid-rapidity \cite{Gelis:2006tb}. The predictions
obtained in all these approaches agree remarkably well. The predicted 
charged particle multiplicity at midrapidity for Pb+Pb is 
$dN_{\rm ch}/dy|_{y=0}\approx 2000\pm 500$. At an initial time 
$\tau_{\rm i}=1\,{\rm GeV}^{-1}=0.2$ fm/c, this corresponds to an 
energy density $\varepsilon_{\rm i}\approx 200$ GeV/fm$^3$, and to 
60 GeV/fm$^3$ at $\tau=0.5$ fm/c. A fully equilibrated quark-gluon 
plasma at these energy densities would have a temperature of 600 MeV
and 450 MeV, respectively.

The higher initial energy density -- about a factor $3$ above that
reached at RHIC (at the same proper time after the onset of the collision) 
-- together with the much higher range of parton-parton
c.~m.~energies $\sqrt{\hat s}$ in the initial state, will give the 
LH-I-C a much larger kinematic range. In particular:
\begin{itemize}
\item 
Jet physics can be probed into the region $E_T > 100$ GeV.
\item 
$b$ and $c$ quarks become plentiful hadronic probes of the medium.
\item
The increased lifetime of the quark-gluon plasma phase, compared to
the final hadronic gas phase, reduces the importance of hadronic 
final-state effects even further for most hard probes, with the 
possible exception of low-mass lepton pairs. 
\end{itemize}
Two remarks qualifying the last statement are in order. Firstly, the
higher initial temperature increases the lifetime of the quark-gluon
plasma phase and, owing to an even more rapid transverse expansion of 
the fireball, helps shorten the final hadronic gas phase. 
Secondly, the increased transverse expansion velocity, compared 
with RHIC collisions, will boost hadronic final-state effects to higher
transverse momentum. This has implications for the $p_T$ distributions
of low-mass lepton pairs and heavy quarkonium states produced, e.~g., 
by late-time recombination.

\section{Hard Probes at LH-I-C}

Hard Probes are Standard Model observables that can be predicted 
perturbatively, with the exception of some infrared sensitive quantities 
that, either, can be determined from other measurements or by means 
of reliable lattice simulations, or are the quantities to be probed.
Example of hard probes are:
\begin{itemize}
\item[$\circ$] High-$p_T$ hadrons.
\item[$\circ$] High-$p_T$ di-hadrons (or $\gamma$+hadron).
\item [$\bullet$] Single jets.
\item[$\circ$] $\gamma$-jet correlations.
\item[$\circ$] Heavy quarkonia ($J/\psi$ and $Upsilon$ states).
\item[$\circ$] High invariant mass lepton pairs.
\item[$\circ$] High-$p_T$ photons.
\item[$\bullet$] $W$ and $Z$ bosons.
\end{itemize}
The items in this list marked by an open circle ($\circ$) are accessible
at both, RHIC and LH-I-C; those marked by a solid bullet ($\bullet$)
are (probably?) only accessible at the higher LH-I-C energies.

It is important to understand in which range of kinematic parameters a 
specific probe can be considered as {\em hard} in this sense. An example
of this issue we have been confronted with at RHIC are single high-$p_T$
hadrons. We now know that pions, but not baryons, can be considered as 
hard probes at transverse momenta above $2-3$ GeV/c. For baryons one
needs to go to $p_T$ in excess of 6 GeV/c for the perturbatively 
calculable fragmentation process to dominate over other modes of production.
It is unclear at present, where the boundaries of the hard probe domain
lie for single hadrons at LH-I-C. We still do not know in which 
kinematic range charmonium states serve as hard probes of the matter
produced at RHIC.

It is equally important to understand how sensitively the extraction of 
physics from a hard probe depends on a detailed understanding (and 
sufficiently realistic modeling) of bulk matter properties. An example
is the extraction of the jet quenching parameter $\hat q$ from single 
inclusive hadron spectra at RHIC. We have come to understand -- and 
the evening discussions at this conference have greatly contributed to
this understanding -- that the different values for $\hat q$ obtained 
by different groups (which range over at least one order of magnitude!)
are not due to a basic lack of understanding of the fundamental process
of parton energy loss, but due to the very much different assumptions 
made about the evolution of the medium and the procedures used to relate
the quenching calculation to the fireball geometry and its evolution.
A transport coefficient, like $\hat q$, can only be considered to have 
a physical meaning if a procedure independent determination from the
comparison with data is achieved.

Returning to the improved accessibility of hard probes at LH-I-C,
it is useful to review some predicted yields \cite{Vogt:2004hc}:
\begin{itemize}
\item
Overall charm and bottom production is predicted to increase by a 
factor 10 and 100, respectively, compared with RHIC.
\item
About 100 $c$-quark pairs and 5 $b$-quark pairs are predicted to be
created in a central Pb+Pb collision at the top LH-I-C energy.
\item
At design luminosity and top energy, one expects about 20 jets with 
total $E_T > 100$ GeV per second from Pb+Pb collisions at LH-I-C.
\item
Under the same conditions, Pb+Pb collisions will yield about one 
$W$-boson per second and 1 $Z$-boson every three seconds.
\end{itemize}
At the same time, $dN/dy$ is expected to increase only by a factor 3;
$dE_T/dy$ by a factor 5. Thus, hard probes will strongly grow in
abundance relative to soft particles.

It is useful to ask whether we are ready to make well founded
predictions for hard probes at LH-I-C energies. The ground work for such
predictions has been laid by several ``yellow book'' publications covering
nuclear effects on parton distributions functions \cite{Accardi:2004be},
jets \cite{Accardi:2004gp}, heavy flavors \cite{Bedjidian:2004gd}, and
electromagnetic probes \cite{Arleo:2004gn}. Quite a bit has been learned
in the RHIC physics program about hard probes since then. This experience
allows us to address the question for which probes we currently have a 
coherent and reasonably complete theoretical framework. Among these are:
\begin{itemize}
\item Single high-$p_T$ photons and hadrons.
\item Single jets.
\item photon-jet correlations.
\item High-invariant mass lepton pairs.
\end{itemize}
The consistency and completeness of the existing framework is much less
clear for di-hadrons at high $p_T$, both, within a single jet and in 
opposite-side jets, and for intermediate $p_T$ photons. On the other
hand, a consistent and comprehensive theoretical framework for heavy 
quarkonia is still elusive.

\section{Case Study: Jets and High-$p_T$ Partons}

Instead of an attempt to provide a comprehensive review of the status 
of all hard probes, which could not be adequately done in this format,
it is instructive to consider one example in more detail: high-$p_T$
hadrons. The perturbative QCD framework for this probe is based on the 
concept of factorization. The differential cross section for inclusive
di-hadrons in opposite jets created by a hard parton-parton scattering 
event of virtuality $Q^2$ is written as
\begin{equation}
\sum_X \frac{d\sigma_{AA'\to hh'+X}}{dQ^2}
= \sum_{p,p'} F^{(1)}_{A\to p}F^{(2)}_{A'\to p'}
   \bigotimes   \sum_{{\bar p},{\bar p}'} 
   \frac{d\sigma_{pp'\to{\bar p}{\bar p}'}}{dQ^2}
   \bigotimes
   {\tilde D}^{(1)}_{{\bar p}\to h} {\tilde D}^{(2)}_{{\bar p}'\to h'},
\label{eq:fact}
\end{equation}
where the $\bigotimes$ symbols indicate convolution over the appropriate
kinematic variables. ${\tilde D}_{{\bar p}\to h}(z)$ is the fragmentation 
function of final-state parton $\bar p$ in the presence of the medium, which 
differs from the vacuum fragmentation function $D(z)$. In the framework of the
twist expansion of perturbative QCD, ${\tilde D}(z)$ can be expressed in terms
of $D(z)$ and a gluon correlator in the medium traversed by the final-state
parton \cite{Guo:2000nz,Wang:2001if} or, equivalently, by the energy loss 
parameter $\hat q$ \cite{Baier:1996sk}. Details of this formulation and its 
application to high-$p_T$ hadron production in nuclear collisions can be 
found in many publications (see e.~g.~\cite{Gyulassy:2003mc} for a review).

What is worth emphasizing here is that this general theoretical framework 
is not enough. The extraction of the parameter $\hat q$ characterizing the
stopping power of the medium requires a detailed modeling of the reaction 
geometry (distribution of scattering vertices, initial density distribution, 
partonic path lengths, longitudinal and transverse expansion, etc.). It is
clear that the value of $\hat q$ extracted from the data is correlated with 
assumptions about the path length $L$ and expansion pattern. As a result 
of these additional assumptions, the extracted values presently range 
widely from ${\hat q} = 0.5 - 15$  GeV$^2$/fm.

It is also important to recognize that the prediction of the nuclear 
suppression factor $R_{AA}$ for single inclusive hadrons containing heavy 
quarks has failed. The observed suppression of $D$-mesons in Au+Au 
collisions at RHIC can barely be described in this pQCD-based framework
by considering the additional energy loss due to elastic collisions 
between the energetic final-state parton and thermal partons in the medium
\cite{Wicks:2005gt}, but {\em only} if the inevitable feed-down from 
decaying $B$-mesons is ignored. Similarly, the large contribution to 
hadron (especially, baryon) production in the intermediate $p_T$ range 
at RHIC was unexpected and is not describable within the framework of 
eq.~(\ref{eq:fact}).

Keeping these chastening facts in mind, let us consider the predictions
of inclusive hadron suppression ($R_{AA}$) at LHC energies. These vary
considerably. For example, Vitev et al. predict that $R_{AA}$ in central 
collisions of heavy nuclei at the LHC rises from about 0.1 at $p_T\sim 10$ 
GeV/c to about 0.4 for $p_T > 100$ GeV/c \cite{Vitev:2002pf,Vitev:2006uc}. 
On the other hand, Eskola et al. predict a value of $R_{AA}\approx 0.15$
that stays roughly constant over this momentum range \cite{Eskola:2004cr}.
As Loizides has analyzed in some detail, the variation between these 
predictions arises from the different treatment of the effective path-length 
distributions for partons with various initial energies. In single inclusive 
measurements only the most energetic partons can exploit the long paths 
associated with scattering vertices in the center of the fireball. 

What applies to single hadrons does not hold for two-parton coincidences. 
Because such measurements effectively fix the vertex to be near the surface 
region on the near (trigger) side, they allow for the exploration of long 
path lengths of opposite-side partons and thus facilitate a more detailed 
study of the
jet quenching mechanism and determination of the energy loss parameter
\cite{Renk:2006ja,Renk:2006pw} even for moderate $p_T\sim 25$ GeV/c. 
Other observables that will be exploited at LH-I-C to study the 
properties of the produced medium are the changes in heavy-to-light 
meson ratios as a function of $p_T$, which probe the quark mass and color
charge dependence of the partonic energy loss \cite{Armesto:2005iq}, and 
$\gamma$-hadron coincidences, which permit the ``tagging'' of the $p_T$
of the scattered quark \cite{Renk:2006qg}.

The much extended kinematic range of parton-parton scattering at  
LH-I-C, compared with RHIC, will make it possible to study the medium 
modified fragmentation function ${\tilde D}(z)$ rather than just the 
change in the distribution of leading hadrons. The observation of 
entire jets, instead of single hadrons, on an event-by-event basis 
at LH-I-C will facilitate the study of medium induced changes in
the jet shape, which are characteristic of the energy loss mechanism
\cite{Salgado:2003rv}. For example, it should be possible to subtract
the underlying soft particle distribution with sufficient accuracy 
to determine ${\tilde D}(z)$ for a 150 GeV jet down to $z\sim 0.02$,
because only a few of the roughly 300 hadrons contained in the jet cone
will have a $p_T > 3$ GeV/c. Finally, the increased transverse flow 
generated by the higher initial pressure of the medium, combined with 
the increased abundance of minijets, will extend the range of quark 
recombination as the dominant mechanism of hadron formation to larger, 
maybe even much larger momenta \cite{Fries:2003fr,Hwa:2006zq}, at least
for baryons.

\section{Other Hard Probes}

Turning to heavy quarks, there exists a solid theoretical framework 
in pQCD for the elementary production of heavy quark pairs, owing to 
the large virtuality scale ($Q^2 = 4m_Q^2$) involved. Much progress
has been made in recent years to develop this framework into the 
ability to make quantitative predictions (see e.~g.~\cite{Baines:2006uw}).
For the more exclusive process of primary quarkonium formation, there 
exist at least two theoretical frameworks, one grounded in a marriage
of perturbative QCD with the nonrelativistic limit of QCD, where color
octet quarkonium states need to be included in the factorization
\cite{Braaten:1994vv}, the other one invoking a heuristic model for 
color ``evaporation'' in the final state \cite{Amundson:1996qr}. The 
agreement of the color octet model with data can be improved by using the 
$k_T$-factorization approach \cite{Kniehl:2006sk}. However, some salient 
predictions of the color octet model (charmonium polarization) have not 
been confirmed (for a recent review, see e.~g.~\cite{Lansberg:2006dh}). 

The situation becomes even more befuddled when one considers the 
interaction of heavy quarkonium states with a QCD medium, for which 
no comprehensive formulation exists at this time. Lattice calculations
have recently succeeded in determining the spectral function of a 
($c\bar c$) pair in various spin-parity channels by means of analytic
continuation of the Euclidean correlation function into Minkowski space
\cite{Asakawa:2003re,Datta:2003ww,Umeda:2005pk}. The surprising result
is that, in contrast to earlier expectations, the $J/\psi$ and $\eta_c$
states survive to temperatures far in excess of $T_c$, at least until
$1.5\, T_c$. This behavior is difficult to reconcile quantitatively 
with the predictions of potential models \cite{Mocsy:2005qw}, which provide
such an excellent description of the vacuum properties of charmonium.

There exists presently no consistent treatment of the interactions of
heavy quarkonium states with a quark-gluon plasma off equilibrium,
which is comparable to the framework developed to describe parton energy
loss. If gluons are the dominant source of parton energy loss in the
medium, one expects them to also contribute to the dissociation of 
heavy quarkonium states as these propagate through the medium
\cite{Grandchamp:2002wp}. The predicted suppression effect at RHIC is
quite substantial, and it becomes very large at LH-I-C 
\cite{Bedjidian:2004gd}. On the other hand, charmonium formation may 
occur within the medium by recombination of independently produced $c$ 
and $\bar c$ quarks \cite{Thews:2000rj}. In the limit where $c$-quarks
thermalize in the medium, this will lead to the statistical emission 
of charmonium at an elevated level dictated by the initial ($c\bar c$) 
production from hard processes \cite{Andronic:2003zv}. Such a scenario 
would result in a substantial enhancement of $J/\psi$ emission at the
LH-I-C. 

In contrast to phenomena related to heavy quarks, electromagnetic
probes of hot and dense matter benefit from a well developed theoretical
framework. This is true across a wide range of processes, from initial
production by hard QCD processes to thermal radiation of photons and 
lepton pairs \cite{Gale:2005zd}, and even for nonequilibrium processes 
associated with the passage of hard partons through the thermalized 
medium \cite{Turbide:2005fk,Turbide:2006mc}. The importance of using 
sophisticated models of the space-time evolution is also increasingly
recognized. The results from RHIC for real and almost-real photons agree 
well with the predictions. This suggests that electromagnetic probes of 
matter are under good theoretical control, and there is little reason to
expect that this will be different at LHC energies. Electromagnetic
probes also assume an increasingly important role in support of other 
hard probes. Charmonium and Upsilon states that decay into lepton pairs
are a well-known example. The tagging of jet energies by direct photons 
\cite{Wang:1996yh} and the probing of the partonic content of the medium 
via jet-to-photon conversion \cite{Fries:2002kt} are other examples, 
which will be important probes at LH-I-C.

\section{Summary and Outlook}

Two years before the start of the LHC heavy ion program, different hard
probes of hot QCD matter are in different stages of development. Generally,
the theory of electromagnetic probes is well developed and a consistent 
theoretical framework for quantitative predictions exists. Probes based
on the interaction of hard partons or jets with the medium still require 
an improved and realistic treatment of evolving fireball geometry including
collective flow. The inelastic interaction between the jet and the medium 
also needs additional conceptual clarifications. For example, what is the 
difference between collisional energy loss, in which the hard parton 
exchanges a virtual gluon with a thermal parton, and radiative energy 
loss, when the radiated gluon is eventually absorbed on a thermal parton?
This example suggests that the treatment of the energy loss of the leading 
parton (the jet initiator) must be imbedded into a complete theory of the 
evolution of the jet inside the medium. The least developed set of hard 
probes are those associated with heavy quarks, where a comprehensive 
theoretical framework for the production and propagation of heavy quarkonia 
in the medium is still missing.

It will be interesting to see whether the matter produced in heavy ion
collisions at the LHC is qualitatively different from that produced in
collisions at RHIC. One important question in this respect is whether 
the QCD plasma produced at RHIC is one with inherently strong coupling
or a less strongly coupled, but turbulent plasma with anomalous transport
coefficients \cite{Asakawa:2006tc}. The former scenario could entail 
significantly different behavior of the medium at early times, when the
temperature is far above $T_c$; the latter would suggest that matter at
LH-I-C looks just like matter at RHIC, only hotter. It remains a 
challenge to theorists to figure out how hard probes can be used to
decide between these two scenarios. Finally, it is worth keeping in 
mind that there may be new surprises waiting at LH-I-C, just
as some key aspects of the RHIC data came as a surprise. In order to 
prepare for the startup of LH-I-C and to separate true surprises 
from physics that should have been anticipated on the basis of what 
we have learned at RHIC, it is important to fill in the mentioned gaps 
in the theoretical framework of hard probes and to make quantitative 
predictions for LHC energies based on state-of-the-art evolution models
of the matter produced in heavy ion collisions.

{\it Acknowledgments:}
This work was supported in part by grants from the U.~S.~Department of
Energy (DE-FG02-05ER41367) and the National Science Foundation 
(NSF-INT-03-35392). I thank the members of the Yukawa Institute for 
Theoretical Physics (YITP), especially Teiji Kunihiro, for their 
hospitality.

\end{document}